\documentclass[sigconf]{acmart}

\AtBeginDocument{%
  }

 \acmConference[ICAIL '25]{}{June 16--20,
2025}{Chicago, IL}
\usepackage{booktabs} 
\usepackage[final]{pdfpages}
\usepackage{amsmath}
\usepackage{xspace}
\usepackage{xcolor}
\usepackage{float}
\usepackage{caption}

\copyrightyear{2025}
\acmYear{2025}
\setcopyright{cc}
\setcctype{by}
\acmConference[ICAIL 2025]{20th International Conference on Artificial Intelligence and Law}{June 16--20, 2025}{Chicago, IL, USA}
\acmBooktitle{20th International Conference on Artificial Intelligence and Law (ICAIL 2025), June 16--20, 2025, Chicago, IL, USA}
\acmPrice{}
\acmDOI{10.1145/3769126.3769199}
\acmISBN{979-8-4007-1939-4/25/06}

\begin{document}

\title{Which Demographic Features Are Relevant for Individual Fairness Evaluation of U.S. Recidivism Risk Assessment Tools?}

%\title{Technical AI Fairness Criteria Might Constitutionally Apply to U.S. Recidivism Risk Assessment Tools, but How and for Whom?}


\author{Tin Trung Nguyen}
\email{tintn@umd.edu}
\affiliation{%
  \institution{University of Maryland}
  \city{College Park}
  \state{Maryland}
  \country{USA}
}

\author{Jiannan Xu}
\email{jiannan@umd.edu}
\affiliation{%
  \institution{University of Maryland}
  \city{College Park}
  \state{Maryland}
  \country{USA}
}

\author{Phuong-Anh Nguyen-Le}
\email{nlpa@umd.edu}
\affiliation{%
  \institution{University of Maryland}
  \city{College Park}
  \state{Maryland}
  \country{USA}
}

\author{Jonathan Lazar}
\email{jlazar@umd.edu}
\affiliation{%
  \institution{University of Maryland}
  \city{College Park}
  \state{Maryland}
  \country{USA}
}

\author{Donald Braman}
\email{dbraman@law.gwu.edu}
\affiliation{%
  \institution{George Washington University}
  \city{Washington}
  \state{DC}
  \country{USA}
}

\author{Hal Daumé III}
\email{hal3@umd.edu}
\affiliation{%
  \institution{University of Maryland}
  \city{College Park}
  \state{Maryland}
  \country{USA}
}

\author{Zubin Jelveh}
\email{zjelveh@umd.edu}
\affiliation{%
  \institution{University of Maryland}
  \city{College Park}
  \state{Maryland}
  \country{USA}
}

%\renewcommand{\shortauthors}{Nguyen et al.}

\begin{abstract}
%Despite its connection to the U.S. constitutional ``Equal Protection" clause thanks to judicial interpretation, 
Despite its constitutional relevance, the technical ``individual fairness'' criterion has not been operationalized in U.S. state or federal statutes/regulations. We conduct a human subjects experiment to address this gap, evaluating which demographic features are relevant for individual fairness evaluation of recidivism risk assessment (RRA) tools. Our analyses conclude that the individual similarity function should consider age and sex, but it should ignore race. 
\end{abstract}






\maketitle

\section{Introduction}
In the U.S. and around the world, AI-assisted decision making tools used in high-stakes contexts like recidivism risk assessment (RRA) for bail/sentencing/parole \cite{brennan2009evaluating}, job application screening \cite{rigotti2024fairness} and loan approval \cite{purificato2023use} have raised bias/fairness concerns. In response, the Machine Learning (ML) community has developed a rich literature in algorithmic fairness metrics \cite{pessach2022review}, with major categories such as group fairness (e.g., equalized positive outcomes across demographic groups) \cite{pedreshi2008discrimination}, individual fairness (similar individuals should get similar outcomes) \cite{dwork2012fairness}, and procedural fairness (perceived permissibility of a feature in the predictive input space) \cite{grgic2018beyond}.

However, technical fairness criteria might be hard to enforce without legal incentives. \citet{nguyen2025how} reviewed primary sources of U.S. state/federal law and found that while three major technical fairness criteria (procedural, group, and individual fairness) can be traced back to judicial interpretation of the U.S. ``Equal Protection'' and ``Due Process'' constitutional clauses, individual fairness has yet been operationalized in specific statutes or regulations. 
To facilitate the development of more concrete legal criteria on individual fairness, we conduct an original human subjects experiment to determine whether individual fairness should consider demographic features like race, sex, or age in the individual similarity function.\footnote{
We store \textbf{Supplementary Materials} (e.g., survey form, data, Python code) in a Google Drive folder: \url{https://drive.google.com/drive/folders/1OHyRIQ21ECEZwD2i1xE_7Ahzhqt9w1IO?usp=drive_link}
}

Our work contributes to the rich literature on humans' perception of algorithmic fairness. 
For example,
\citet{wang2020factors} studied how different demographic groups of participants perceive fairness differently when an AI model makes demographically biased predictions in the participants' favors.
\citet{harrison2020empirical} conducted surveys where participants were asked to compare two models representing trade-offs between fairness-related properties. They observed a marginal preference for equalizing false positives across demographic groups over equalizing accuracy. 
\citet{saxena2019fairness} explored people's attitudes about individual fairness, investigating which definitions people perceive to be the fairest in the context of loan decisions, and whether fairness perceptions change with the addition of sensitive information (i.e., race of the loan applicants). 
\citet{grgic2016case} surveyed 100 Mechanical Turk workers' fairness perception when using different features in an RRA task. 
These studies underscore the complexity of determining relevant fairness criteria when demographic factors are considered.

\section{Methodology -- Human Study}
\label{section-backward}

We design a human subjects experiment to infer whether a feature should be excluded from an individual similarity function based on people's perception of whether the outcomes of each pair, whom the similarity function determines as highly similar, are fair.

Assuming an individual similarity function is good, then the intuition behind individual fairness is that given two individuals with a high similarity score ($S=high$) according to the individual similarity function, the judgment is fair ($J=fair$) if those two individuals receive the same outputs ($O=same$) and unfair ($J=unfair$) if they receive different outputs ($O=diff$). If the similarity score for two individuals is low ($S=low$), individual fairness cannot make any judgment ($J=null$). We formalize these intuitions with the logical and ($\land$) as well as the logical or ($\lor$) as follows:

Assumption: similarity function is good.
\begin{align}
(S = high \, \land \, O=diff) \Rightarrow J = unfair \nonumber 
\end{align}

Note that the equivalent of $A \Rightarrow B$ is $\neg A \lor B$ where $\neg$ is logical negation. We can rewrite the statement above as:
\begin{align}
\neg(S = high \, \land \, O=diff) \lor (J = unfair) \nonumber \\
\iff  S = low \, \lor \, O=same \lor J = unfair  \label{logic2}
\end{align}

Assuming a similarity function is good means that Statement \ref{logic2} must be true. If Statement \ref{logic2} is false, then we know the similarity function is bad. To quantify the ``badness'' of a similarity function, we use the pairs of individuals that the similarity function determines as highly similar ($S=high$). We can conduct a human subjects experiment (human study) where crowdworkers make human judgments ($J=fair$ or $J=unfair$) on pairs of dissimilar recidivism outputs. 
Let $O=same$ if two individuals are classified into the same level of recidivism risk, e.g., both Low, or both High; and $O=diff$ otherwise, e.g., one Low and one High. The treatment variable is the similarity function, i.e., each crowdworker sees a set of features that only one of the individual similarity functions uses. 
A pair with ($S=high \, \land \, O=diff \, \land \, J = fair$) violates Statement \ref{logic2}, incrementing the ``badness'' score of the similarity function by one unit. Based on fairness judgments on the highly different recidivism risk scores in pairs of highly similar defendants, we can compute an overall ``badness'' score of the similarity function. 

Back to the question of how to empirically determine whether a feature should be excluded from the similarity function, we can repeat the backward evaluation procedure above with the same type of similarity function but consider different sets of features.

Once we decide on a simple type of similarity function, we can apply it to a set of features that should certainly be included in the similarity function, e.g., criminal history features, and use this similarity function to run the ``badness'' score estimation procedure above to get a baseline ``badness'' score. Then we can evaluate a new similarity function that has the same form but additionally takes into account a feature in dispute, e.g., race, and repeat the procedure to get a ``badness'' score for this new similarity function. At the final step, if the ``badness'' score of the similarity function with race is (with statistical significance) higher than the baseline similarity function (without any disputed features), we may conclude that based on humans' fairness perception of the recidivism risk assessment task, race should be ignored, i.e., not included, in the individual similarity function when evaluating individual fairness.
\section{Experimental Set-up}
\subsection{Criminal Justice Dataset}
We draw defendant pairs from the Client Legal Utility Engine (CLUE) dataset 
with 4.4 million District Court and 700,000 Circuit Court cases scraped from the Maryland Judiciary Case Search.\footnote{\url{https://casesearch.courts.state.md.us/casesearch/}} 
It contains criminal history features (e.g., arrests and convictions) and demographic features, e.g. race, sex, and age (at last arrest).

We generate a recidivism prediction for everyone who had an arrest between 2016 to 2018. For these defendants, we build features using four years of history (which spans data from 2013 to 2018) and we predict whether they recidivated (re-arrested for a violent offense) within one year following their initial arrest (2017-2019). After data cleaning and record linkage, we obtain at a subset of roughly 25,000 defendants and select highly similar pairs from it. 

Regarding more details of the recidivism prediction step, we develop a simple AI model (random forest classifier\footnote{\url{https://scikit-learn.org/stable/modules/generated/sklearn.ensemble.RandomForestClassifier.html}}) to predict the (violent) recidivism risk in the 2017-2019 period (on a scale from 0 to 1) of defendants in the CLUE dataset. We set the high recidivism risk threshold to be the $62^{th}$ percentile of predicted risk scores across the felony offenders of the 25,000 defendants in our cleaned CLUE dataset. This choice is inspired by the finding from the ``Pretrial Release of Felony Defendants in State Courts'' report by the DOJ (2007)\footnote{\url{https://bjs.ojp.gov/content/pub/pdf/prfdsc.pdf}} that 62\% of felony defendants between 1990-2004 were released before trial, which might imply that roughly 38\% of felony defendants in that period were considered high risk. The ``high risk'' threshold we get with this method is roughly 0.40.

\subsection{Main Experiment}

After carefully considering the trade-offs between the two types of human subjects experiment (within-subjects vs. between-subjects) as detailed by \citet{mackenzie2024human}, we decided to use a between-subjects experiment setting for two reasons. 
First, we want to avoid the learning effect in the within-subjects setting, i.e., what a participant learned from an earlier condition may influence her decisions in a later condition. 
Second, a disadvantage of between-subjects experiments is ``participant disposition'' (some participants might be particularly more meticulous or reckless than others). However, we can mitigate this disadvantage by adding three attention checks throughout the study (to filter out reckless participants) and a financial bonus to incentivize participants to be more meticulous. 

Therefore, we conducted a between-subjects experiment on Qualtrics (with participants recruited from Prolific). To enhance ecological validity and data quality, we apply the following participant filters: based in the U.S., English proficiency, Prolific approval rate of at least $98\%$, and Bachelor's degree. We also provide fairness voting bonuses to incentivize participants to align their fairness evaluation with common sense and optional rationale bonuses to incentivize them to think carefully about each fairness judgment.
Furthermore, we filter out responses from participants who fail two or more attention checks. The median completion time among participants who passed our attention checks is $18.5$ minutes. We pay each participant $\$3.7$, corresponding to $12$ USD per hour. 

\begin{figure}
\centering
    \includegraphics[width=0.5\textwidth, trim = 0 100 0 80]{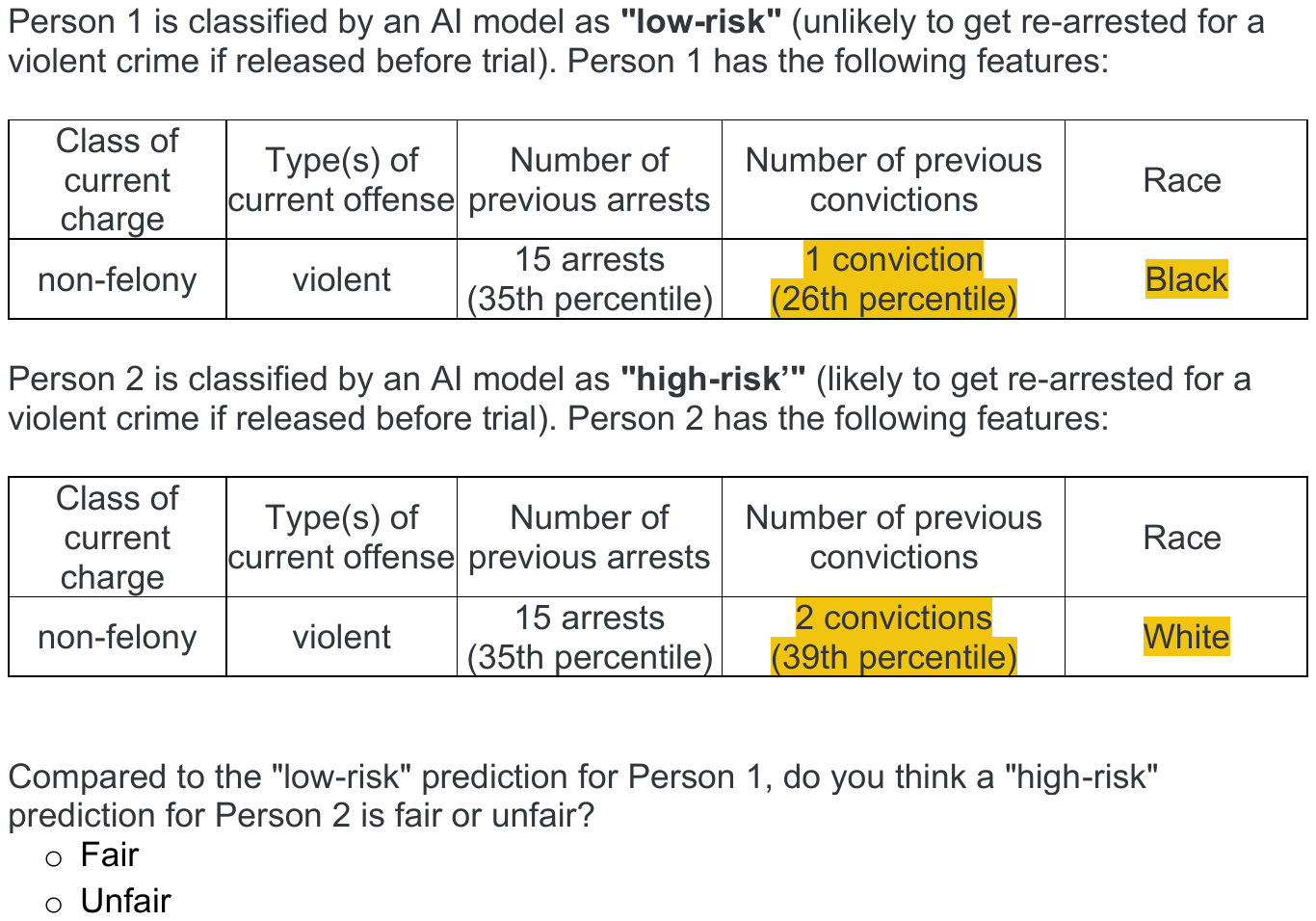}
    \caption{Example interface for the main human study task (Condition: Treatment with Race)}
    \label{interface-race}
\end{figure}

Each participant was randomly assigned to one of the four conditions for the individual fairness evaluation task: one control condition (with only the four criminal history features) and three treatment conditions. One example interface (for the treatment condition with race) is in \autoref{interface-race}. 
Their main task is to decide whether the differing outcomes between two defendants in each pair are fair. 

In each treatment condition, we show the four criminal history features plus one of the three demographic features---race, sex, and age group. The four conditions use data from the same 14 pairs of real-life defendants: we only show a different subset of the defendants' features in each condition. In 10 out of those 14 pairs, the two defendants in the same pair have different values across all the three demographic features (e.g., Black Male, 41-46 years old vs. White Female, 18-22 years old). In the remaining 4 pairs, both defendants in the same pair have the same demographic values.

At the end, to also help operationalize procedural fairness, we have a simple question about which features (criminal history, race, sex, or age) each participant thinks should be used as model inputs.

This human study was approved by the University of Maryland (UMD)'s Institutional Review Board (IRB), project number 2120460.
\section{Experimental Results}

\subsection{Brief result on procedural fairness}

While most people (91.5\%) think that criminal history features should be used as input features, only 35.5\%, 17.5\%, and 6.0\% of participants think that age, sex, and race should be used, respectively.
We move on to main experimental results on individual fairness.

\subsection{Quantitative analysis: no significant fairness judgement differences across conditions}
\label{quantitative}

\begin{table}
\centering
\begin{tabular}{cccc}
\toprule
        & Ratings & Fair & ``Fair'' Rating Rate (\%) \\
\midrule
Control & 854        & 461  & 53.98          \\
Race    & 798        & 446  & 55.89          \\ 
Age     & 840        & 491  & 58.45          \\
Sex     & 784        & 425  & 54.2           \\
\bottomrule
\end{tabular}
\caption{The ``Fair'' rating rate per condition. Each of $N=234$ participants who pass all attention checks gives $14$ ratings.}
\label{tab:exp_results}
\end{table}

The ``Fair" rating rate from the \{Race, Age, Sex\} conditions is slightly higher than what the control condition yields, as in \autoref{tab:exp_results}. 
However, it might be the case that the differences were caused by inherent random noises. We use chi-square tests to check which condition has a significantly higher rate of ``Fair'' ratings. Recalling that we have four conditions (Control, Race, Age, Sex) in total, we test two competing hypotheses for each pair of experimental conditions as follows where $(C_i, C_j)$ denotes a pair of designs. 
\begin{itemize}
    \item $H_0$ (Null Hypothesis): There is no significant difference in fair rating rates in $(C_i, C_j)$, $i \neq j \in \{Control, Race, Age, Sex\}$. 
    \item $H_A$ (Alternative Hypothesis): There is a significant difference in fair rating rates in $(C_i, C_j)$, $i \neq j \in \{Control, Race, Age, Sex\}$. 
\end{itemize}
Since we use multiple simultaneous statistical tests, we use Bonferroni correction to counteract the multiple comparisons problem. 
\autoref{tab:chisq_tests} shows that the chi-square statistics for several pairs of conditions are not statistically significant under Bonferroni correction. 

%\begin{table}[H]
\begin{table}
\centering
\resizebox{0.5\textwidth}{!}{\begin{tabular}{cccc}
\toprule
                & Chi-square Statistics & P-values & P-values (Bonferroni Correction) \\ 
\midrule
(Control, Race) & 0.6068                & 0.4360   & 1.0                              \\
(Control, Age)  & 3.4391                & 0.0637  & 0.3820                           \\
(Control, Sex)  & 0.0086                & 0.9263   & 1.0                          \\
(Race, Age)     & 1.0978                & 0.2948   & 1.0                          \\
(Race, Sex)     & 0.4514                & 0.5017   & 1.0                        \\
(Age, Sex)      & 2.9692               & 0.0849 &  0.5092                            \\ 
\bottomrule
\end{tabular}}
\caption{The chi-square tests for each pair of conditions}
\label{tab:chisq_tests}
\end{table}

More statistical tests, including OLS Regression, Logistic Regression, and Mixed Effect Linear Regression, also showed no significant difference across our four experimental conditions. Therefore, rigorous quantitative analysis shows no evidence that including a demographic feature (race, sex, or age) would worsen people's individual fairness judgment (by making them more likely to select ```Fair''' even though similar defendants got opposite outcomes). However, this result suggests another potential effect that the multiple-choice questions alone may not capture. Some participants might have internalized certain social norms to not consider certain demographic features (e.g., race) at all when making fairness judgments. We conduct qualitative analysis to investigate whether the quantitative result is because people simply refuse to consider each demographic feature or because each demographic feature, when thoroughly considered, does not worsen people's individual fairness judgment.

\subsection{Qualitative analysis: Race and Sex are often ignored, but Age impacts fairness judgment.} 
\label{qualitative}

We qualitatively code the optional rationales after each ``Fair'' vs. ``Unfair'' vote in the treatment conditions, to check whether the lack of quantitative evidence (that demographic features worsen fairness judgement) is simply because participants ignore each respective demographic feature, or due to more nuanced considerations.

 %\begin{table*}[t]
\begin{table*}
    \centering
    \resizebox{1.0\textwidth}{!}{\begin{tabular}{cccccc}
    \toprule
        Treatment feature  & Rationales with keyword  & (Fair $\times$ Impact) & (Fair $\times$ No Impact)   & (Unfair $\times$ Impact) & (Unfair $\times$ No Impact) \\
    \midrule
        Race & 13 rationales & 0\% & 0\% & 15.4\% & 84.6\% \\ 
        Sex & 21 rationales & 14.3\% & 14.3\% & 19.0\% & 52.4\% \\ 
        Age (group) & 66 rationales & 62.1\% & 6.1\% & 16.7\% & 15.1\% \\ 
        \bottomrule
    \end{tabular}}
   \caption{Statistics on the Qualitative Clusters of Optional Rationales (the Impact vs. No Impact dimension was coded by us)}
    \label{qualitative_table}
\end{table*}

%\kem{is it a bug that table 3 is coming before table 2?} 

Method-wise, we select the rationales that contain one of the three demographic keywords (``race'', ``sex'', ``age''). If there are duplicated rationales from the same participant for several questions, we keep the first rationale. As each rationale has already been associated with two dimensions -- a demographic feature and a (Fair/Unfair) multiple-choice response, we manually label the rationales along a third dimension: Does Treatment Feature impacts Perceived Fairness? (``Impact'' means that the rationale uses the demographic feature to justify their Fair/Unfair response; ``No Impact'' means the rationale mentions the demographic feature, but then either ignores it or explains that the feature does not influence in their fairness evaluation).

Clustering the coded rationales along the three dimensions, we get two qualitative findings from \autoref{qualitative_table}. 

First, for the two Race and Sex treatment features, the (Unfair $\times$ No Impact) cluster is larger than the other three clusters by a great margin. For these two treatment features, our qualitative analysis agrees with the quantitative finding that showing demographic features does not affect fairness perception (and thus the pairwise distance function) because the participants often do not take these demographic features into account (even after they see and cite the features), e.g., 1. ``Both are almost equal. Race shouldn't be a factor in deciding.'' and 2. ``Both should be high risk, the sex of the individual should not warrant such a drastic difference.''

Second, for Age, the biggest cluster is (Fair $\times$ Impact), which is the most seemingly concerning cluster because rationales here show that the demographic feature worsens the quality of the pairwise individual similarity function, by making participants less likely to recognize that the two defendants in the same pair are highly similar and therefore less likely to select the ``Unfair'' vote. The main recurring theme in this cluster is that it's fair for a younger person to receive a higher risk prediction, e.g., ``The younger person may not be as mature as the older one, but also they have a lot more time in years cause they're young to commit more crimes.''
For the next biggest cluster, a frequent argument in the (Unfair $\times$ Impact) cluster is that being in an older age group means a defendant got enough time to rehabilitate, so their slightly worse criminal record is not enough to justify their higher predicted risk, e.g., ``Both should have the same risk. Person 1 does not have a violent conviction. Though person 2 has a prior conviction their age must be considered, this could have been from over 20 years ago.''

It is tempting to conclude that Age worsens the quality of the pairwise individual similarity function more than the other two demographic features because Age is the only feature with a majority (Fair $\times$ Impact) cluster. However, combining this observation and the lack of evidence for cross-condition significant difference,  
we conclude that Age does not necessarily make more people more likely to choose the ``Fair'' judgment, but this feature prompts people who choose the ``Fair'' judgment to think out loud more (by writing a detailed rationale citing age) to justify their choice -- a constructive impact to enable more transparent decisions.

\section{Discussion: Which Demographic Features Should the Individual Similarity Function Ignore?}

\paragraph{Age.}
Considering the aforementioned constructive impact of age, we further examine four more arguments that favor including this feature in the pairwise individual similarity function. 

First, age belongs to the ``rational basis'' category \cite{1976massachusetts},\footnote{Massachusetts Bd. of Retirement v. Murgia (1976): ``We [U.S. Supreme Court] turn then to examine this state [age-based] classification under the rational basis standard.''} which includes the least judicially scrutinized features in demographic classification by the government, so it is more likely than race or sex to be justifiable as part of individual fairness evaluation. 

Second, due to the continuous nature of age, a small difference in age, if close to a legal threshold, can yield significant differences in the standard of proof a person is subject to and, therefore, to the conviction outcome. For example, while American adults have a constitutional right to a jury trial, the U.S. Supreme Court ruled in McKeiver v. Pennsylvania (1971) that this constitutional right does not extend to juvenile defendants \cite{1971mckeiver}, which might imply that the standard of proof for an adult versus a juvenile might differ due to the difference in the number of fact finders between a bench trial and a jury trial. A more specific example regarding the age-based difference in the standard of proof is the case In re Mitchell P (1978), where the California Supreme Court ruled that ``uncorroborated testimony of an accomplice'' is not forbidden as evidence in California juvenile courts \cite{1978michell}, even though California criminal courts (for adults) forbid such evidence, as pointed out by \citet{mcgoldrick1978juvenile}. These subtle differences in standard of proof may make two highly similar-aged defendants (e.g., one old juvenile and one early adult) receive very different criminal history features (such as past convictions). If a pairwise individual similarity function does not take into account a defendant's age but only the criminal history features, this function might assume that those two highly similar-aged defendants are very different in the input space, and thus make no (un-)fairness alerts on their disparate risk scores. 

Third, certain age groups have higher average violence prevalence. As it is reasonable to assume that people with similar violence prevalence should be more likely to get similar (violent) recidivism predictions, age might be a reasonable proxy for violence prevalence. For example, \citet{fahlgren2020age} empirically show that younger individuals are more likely to commit a violent act.

Fourth, some judges give more leniency to younger offenders \citep{stevenson2022algorithmic}. Therefore, including age in the individual similarity function might uncover cases where leniency discretion is applied inconsistently by a judge across similar-aged defendants.

Therefore, the individual similarity function should include age.

\paragraph{Sex.}
The main arguments for including sex, typically an ``intermediate scrutiny'' feature \cite{1976craig},\footnote{Craig v. Boren (1976): ``classifications by gender must serve important governmental objectives and must be substantially related to achievement of those objectives.''}
into the individual similarity function are that first, our empirical study shows that including sex does not significantly worsen people's fairness judgement, and second, the violence prevalence argument for age also applies to sex (males are often more capable of violence than females). Therefore, the individual similarity function should also include sex.

\paragraph{Race.}
Decided by the U.S. Supreme Court as a ``strict scrutiny'' feature \cite{1944korematsu},\footnote{Korematsu v. United States (1944): “all legal restrictions which curtail the civil rights of a single racial group are immediately suspect. [...] courts must subject them to the most rigid scrutiny.”}
race is a nuanced case. On the one hand, past studies have shown correlations between race and several features that are related to violence prevalence. For example, \citet{williams2007understanding} show that among urban juveniles, race has a significant correlation with `initiation of [under-aged] alcohol use' while sex does not. Between two minority groups of juveniles (Black vs. Hispanic), \citet{reingle2012racial} find significantly higher longitudinal level of aggression among Black juveniles. If one believes that defendants with similar prevalence of violence should get similar recidivism prediction, then one might argue that race should be used in the individual similarity function as one of the proxies to estimate similarity in terms of violence prevalence.

On the other hand, \citet{mcnulty2003explaining} find that the correlation between race and `involvement in violence' can be explained by the ``community disadvantage'' variable when comparing Black adolescents with other racial groups of adolescents. Therefore, one can make a normative argument that even if race correlates with violence prevalence, this correlation is an artifact of historical discrimination and therefore should not be used in an individual fairness evaluation pipeline.
Consider additionally a hypothetical: Two identical defendants A and B, with all features being identical (same criminal history, same sex, same age) except race, get two opposite RRA predictions (``high risk'' for the Black defendant and ``low risk'' for the White defendant). We assume that normatively, almost everyone will consider this instance as evidence of an (individually) unfair, maybe even ``racist'', algorithm. This normative assumption might not necessarily hold for sex or age as one can still argue that all else equal, assigning different risk scores between a man and a woman, or between an 80-year-old and a 20-year-old, is not necessarily evidence of algorithmic unfairness.
If our pairwise individual similarity function ignores race, it will certainly identify A and B as a highly similar pair (with zero distance), thus counting their opposite outcomes as evidence of individual unfairness. However, if our individual similarity function incorporates race, there is a chance that the function might compare the non-zero distance between A and B with all other pairwise distances and decide that this distance is not small enough to consider A and B as highly similar individuals. In that case, the opposite outcomes of A and B will not count towards the final individual fairness evaluation.
This hypothetical illustrates a key weakness of using race in the individual similarity function, so this function should ignore race.

\paragraph{Conclusion}
We find no statistical evidence that race, sex, or age worsens laypeople's individual fairness perception in the RRA context, but qualitative analysis shows that age does impact and make their fairness judgement more nuanced. Our follow-up legal and policy analyses conclude that individual fairness evaluation of RRA tools should consider age and sex, but it should ignore race.                                                                                                                                                                                                                                
%\section*{Acknowledgement}
\begin{acks}
This study is supported by the National Institute of Justice's Graduate Research Fellowship 2023 (Award 15PNIJ-23-GG-01932-RESS).
\end{acks}

\bibliographystyle{ACM-Reference-Format}
\bibliography{sample-base}


\begin{thebibliography}{24}

%%% the Web address.

\ifx \showCODEN    \undefined \def \showCODEN     #1{\unskip}     \fi
\ifx \showDOI      \undefined \def \showDOI       #1{#1}\fi
\ifx \showISBNx    \undefined \def \showISBNx     #1{\unskip}     \fi
\ifx \showISBNxiii \undefined \def \showISBNxiii  #1{\unskip}     \fi
\ifx \showISSN     \undefined \def \showISSN      #1{\unskip}     \fi
\ifx \showLCCN     \undefined \def \showLCCN      #1{\unskip}     \fi
\ifx \shownote     \undefined \def \shownote      #1{#1}          \fi
\ifx \showarticletitle \undefined \def \showarticletitle #1{#1}   \fi
\ifx \showURL      \undefined \def \showURL       {\relax}        \fi
% invisible to TeX
\providecommand\bibfield[2]{#2}
\providecommand\bibinfo[2]{#2}
\providecommand\natexlab[1]{#1}
\providecommand\showeprint[2][]{arXiv:#2}

\bibitem[Brennan et~al\mbox{.}(2009)]%
        {brennan2009evaluating}
\bibfield{author}{\bibinfo{person}{Tim Brennan}, \bibinfo{person}{William Dieterich}, {and} \bibinfo{person}{Beate Ehret}.} \bibinfo{year}{2009}\natexlab{}.
\newblock \showarticletitle{Evaluating the predictive validity of the COMPAS risk and needs assessment system}.
\newblock \bibinfo{journal}{\emph{Criminal Justice and Behavior}} \bibinfo{volume}{36}, \bibinfo{number}{1} (\bibinfo{year}{2009}), \bibinfo{pages}{21--40}.
\newblock


\bibitem[{California Supreme Court}(1978)]%
        {1978michell}
\bibfield{author}{\bibinfo{person}{{California Supreme Court}}.} \bibinfo{year}{1978}\natexlab{}.
\newblock \showarticletitle{In re Mitchell P}.
\newblock  \bibinfo{number}{{22 Cal.3d 946, 151 Cal. Rptr. 330, 587 P.2d 1144 (Cal. 1978)}} (\bibinfo{year}{1978}).
\newblock


\bibitem[Dwork et~al\mbox{.}(2012)]%
        {dwork2012fairness}
\bibfield{author}{\bibinfo{person}{Cynthia Dwork}, \bibinfo{person}{Moritz Hardt}, \bibinfo{person}{Toniann Pitassi}, \bibinfo{person}{Omer Reingold}, {and} \bibinfo{person}{Richard Zemel}.} \bibinfo{year}{2012}\natexlab{}.
\newblock \showarticletitle{Fairness through awareness}. In \bibinfo{booktitle}{\emph{Proceedings of ITCS 2012}}. \bibinfo{pages}{214--226}.
\newblock


\bibitem[Fahlgren et~al\mbox{.}(2020)]%
        {fahlgren2020age}
\bibfield{author}{\bibinfo{person}{Martha~K Fahlgren}, \bibinfo{person}{Evan~M Kleiman}, \bibinfo{person}{Alexander~A Puhalla}, {and} \bibinfo{person}{Michael~S McCloskey}.} \bibinfo{year}{2020}\natexlab{}.
\newblock \showarticletitle{Age and gender effects in recent violence perpetration}.
\newblock \bibinfo{journal}{\emph{Journal of Interpersonal Violence}} \bibinfo{volume}{35}, \bibinfo{number}{17-18} (\bibinfo{year}{2020}), \bibinfo{pages}{3513--3529}.
\newblock


\bibitem[Grgic-Hlaca et~al\mbox{.}(2016)]%
        {grgic2016case}
\bibfield{author}{\bibinfo{person}{Nina Grgic-Hlaca}, \bibinfo{person}{Muhammad~Bilal Zafar}, \bibinfo{person}{Krishna~P Gummadi}, {and} \bibinfo{person}{Adrian Weller}.} \bibinfo{year}{2016}\natexlab{}.
\newblock \showarticletitle{The case for process fairness in learning: Feature selection for fair decision making}. In \bibinfo{booktitle}{\emph{NIPS Symposium on Machine Learning and the Law}}, Vol.~\bibinfo{volume}{1}. Barcelona, Spain, \bibinfo{pages}{11}.
\newblock


\bibitem[Grgi{\'c}-Hla{\v{c}}a et~al\mbox{.}(2018)]%
        {grgic2018beyond}
\bibfield{author}{\bibinfo{person}{Nina Grgi{\'c}-Hla{\v{c}}a}, \bibinfo{person}{Muhammad~Bilal Zafar}, \bibinfo{person}{Krishna~P Gummadi}, {and} \bibinfo{person}{Adrian Weller}.} \bibinfo{year}{2018}\natexlab{}.
\newblock \showarticletitle{Beyond distributive fairness in algorithmic decision making: Feature selection for procedurally fair learning}. In \bibinfo{booktitle}{\emph{Proceedings of the AAAI Conference on Artificial Intelligence}}, Vol.~\bibinfo{volume}{32}.
\newblock


\bibitem[Harrison et~al\mbox{.}(2020)]%
        {harrison2020empirical}
\bibfield{author}{\bibinfo{person}{Galen Harrison}, \bibinfo{person}{Julia Hanson}, \bibinfo{person}{Christine Jacinto}, \bibinfo{person}{Julio Ramirez}, {and} \bibinfo{person}{Blase Ur}.} \bibinfo{year}{2020}\natexlab{}.
\newblock \showarticletitle{An empirical study on the perceived fairness of realistic, imperfect machine learning models}. In \bibinfo{booktitle}{\emph{Proceedings of FAccT (FAT) 2020}}. \bibinfo{pages}{392--402}.
\newblock


\bibitem[MacKenzie(2024)]%
        {mackenzie2024human}
\bibfield{author}{\bibinfo{person}{I.~Scott MacKenzie}.} \bibinfo{year}{2024}\natexlab{}.
\newblock \showarticletitle{Human-Computer Interaction: An empirical research perspective. Chapter 5.10}.
\newblock  (\bibinfo{year}{2024}).
\newblock


\bibitem[McGoldrick~Jr(1978)]%
        {mcgoldrick1978juvenile}
\bibfield{author}{\bibinfo{person}{James~M McGoldrick~Jr}.} \bibinfo{year}{1978}\natexlab{}.
\newblock \showarticletitle{Juvenile Justice and the Equal Protection Clause: First Class, Tourist, or Luxury Coach}.
\newblock \bibinfo{journal}{\emph{Pepp. L. Rev.}}  \bibinfo{volume}{6} (\bibinfo{year}{1978}), \bibinfo{pages}{697}.
\newblock


\bibitem[McNulty and Bellair(2003)]%
        {mcnulty2003explaining}
\bibfield{author}{\bibinfo{person}{Thomas~L McNulty} {and} \bibinfo{person}{Paul~E Bellair}.} \bibinfo{year}{2003}\natexlab{}.
\newblock \showarticletitle{Explaining racial and ethnic differences in serious adolescent violent behavior}.
\newblock \bibinfo{journal}{\emph{Criminology}} \bibinfo{volume}{41}, \bibinfo{number}{3} (\bibinfo{year}{2003}), \bibinfo{pages}{709--747}.
\newblock


\bibitem[Nguyen et~al\mbox{.}(2025)]%
        {nguyen2025how}
\bibfield{author}{\bibinfo{person}{Tin Nguyen}, \bibinfo{person}{Jiannan Xu}, \bibinfo{person}{Phuong-Anh Nguyen-Le}, \bibinfo{person}{Jonathan Lazar}, \bibinfo{person}{Donald Braman}, \bibinfo{person}{Hal Daumé}, {and} \bibinfo{person}{Zubin Jelveh}.} \bibinfo{year}{2025}\natexlab{}.
\newblock \showarticletitle{How May U.S. Courts Scrutinize Their Recidivism Risk Assessment Tools? Contextualizing AI Fairness Criteria on a Judicial Scrutiny-based Framework}.
\newblock \bibinfo{journal}{\emph{SSRN preprint}} (\bibinfo{year}{2025}).
\newblock
\urldef\tempurl%
\url{https://papers.ssrn.com/sol3/papers.cfm?abstract_id=5242075}
\showURL{%
\tempurl}


\bibitem[Pedreshi et~al\mbox{.}(2008)]%
        {pedreshi2008discrimination}
\bibfield{author}{\bibinfo{person}{Dino Pedreshi}, \bibinfo{person}{Salvatore Ruggieri}, {and} \bibinfo{person}{Franco Turini}.} \bibinfo{year}{2008}\natexlab{}.
\newblock \showarticletitle{Discrimination-aware data mining}. In \bibinfo{booktitle}{\emph{Proceedings of SIGKDD 2008}}. \bibinfo{pages}{560--568}.
\newblock


\bibitem[Pessach and Shmueli(2022)]%
        {pessach2022review}
\bibfield{author}{\bibinfo{person}{Dana Pessach} {and} \bibinfo{person}{Erez Shmueli}.} \bibinfo{year}{2022}\natexlab{}.
\newblock \showarticletitle{A review on fairness in machine learning}.
\newblock \bibinfo{journal}{\emph{ACM Computing Surveys (CSUR)}} \bibinfo{volume}{55}, \bibinfo{number}{3} (\bibinfo{year}{2022}), \bibinfo{pages}{1--44}.
\newblock


\bibitem[Purificato et~al\mbox{.}(2023)]%
        {purificato2023use}
\bibfield{author}{\bibinfo{person}{Erasmo Purificato}, \bibinfo{person}{Flavio Lorenzo}, \bibinfo{person}{Francesca Fallucchi}, {and} \bibinfo{person}{Ernesto~William De~Luca}.} \bibinfo{year}{2023}\natexlab{}.
\newblock \showarticletitle{The use of responsible artificial intelligence techniques in the context of loan approval processes}.
\newblock \bibinfo{journal}{\emph{International Journal of Human--Computer Interaction}} \bibinfo{volume}{39}, \bibinfo{number}{7} (\bibinfo{year}{2023}), \bibinfo{pages}{1543--1562}.
\newblock


\bibitem[Reingle et~al\mbox{.}(2012)]%
        {reingle2012racial}
\bibfield{author}{\bibinfo{person}{Jennifer~M Reingle}, \bibinfo{person}{Mildred~M Maldonado-Molina}, \bibinfo{person}{Wesley~G Jennings}, {and} \bibinfo{person}{Kelli~A Komro}.} \bibinfo{year}{2012}\natexlab{}.
\newblock \showarticletitle{Racial/ethnic differences in trajectories of aggression in a longitudinal sample of high-risk, urban youth}.
\newblock \bibinfo{journal}{\emph{Journal of Adolescent Health}} \bibinfo{volume}{51}, \bibinfo{number}{1} (\bibinfo{year}{2012}), \bibinfo{pages}{45--52}.
\newblock


\bibitem[Rigotti and Fosch-Villaronga(2024)]%
        {rigotti2024fairness}
\bibfield{author}{\bibinfo{person}{Carlotta Rigotti} {and} \bibinfo{person}{Eduard Fosch-Villaronga}.} \bibinfo{year}{2024}\natexlab{}.
\newblock \showarticletitle{Fairness, AI \& recruitment}.
\newblock \bibinfo{journal}{\emph{Computer Law \& Security Review}}  \bibinfo{volume}{53} (\bibinfo{year}{2024}), \bibinfo{pages}{105966}.
\newblock


\bibitem[Saxena et~al\mbox{.}(2019)]%
        {saxena2019fairness}
\bibfield{author}{\bibinfo{person}{Nripsuta~Ani Saxena}, \bibinfo{person}{Karen Huang}, \bibinfo{person}{Evan DeFilippis}, \bibinfo{person}{Goran Radanovic}, \bibinfo{person}{David~C Parkes}, {and} \bibinfo{person}{Yang Liu}.} \bibinfo{year}{2019}\natexlab{}.
\newblock \showarticletitle{How do fairness definitions fare? Examining public attitudes towards algorithmic definitions of fairness}. In \bibinfo{booktitle}{\emph{Proceedings of AIES 2019}}. \bibinfo{pages}{99--106}.
\newblock


\bibitem[Stevenson and Doleac(2022)]%
        {stevenson2022algorithmic}
\bibfield{author}{\bibinfo{person}{Megan~T Stevenson} {and} \bibinfo{person}{Jennifer~L Doleac}.} \bibinfo{year}{2022}\natexlab{}.
\newblock \showarticletitle{Algorithmic risk assessment in the hands of humans}.
\newblock \bibinfo{journal}{\emph{Available at SSRN 3489440}} (\bibinfo{year}{2022}).
\newblock


\bibitem[{U.S. Supreme Court}(1944)]%
        {1944korematsu}
\bibfield{author}{\bibinfo{person}{{U.S. Supreme Court}}.} \bibinfo{year}{1944}\natexlab{}.
\newblock \showarticletitle{Korematsu v. United States}.
\newblock  \bibinfo{number}{323 U.S. 214} (\bibinfo{year}{1944}).
\newblock


\bibitem[{U.S. Supreme Court}(1971)]%
        {1971mckeiver}
\bibfield{author}{\bibinfo{person}{{U.S. Supreme Court}}.} \bibinfo{year}{1971}\natexlab{}.
\newblock \showarticletitle{McKeiver v. Pennsylvania}.
\newblock  \bibinfo{number}{{403 U.S. 528}} (\bibinfo{year}{1971}).
\newblock


\bibitem[{U.S. Supreme Court}(1976a)]%
        {1976craig}
\bibfield{author}{\bibinfo{person}{{U.S. Supreme Court}}.} \bibinfo{year}{1976}\natexlab{a}.
\newblock \showarticletitle{Craig v. Boren}.
\newblock  \bibinfo{number}{429 U.S. 190} (\bibinfo{year}{1976}).
\newblock


\bibitem[{U.S. Supreme Court}(1976b)]%
        {1976massachusetts}
\bibfield{author}{\bibinfo{person}{{U.S. Supreme Court}}.} \bibinfo{year}{1976}\natexlab{b}.
\newblock \showarticletitle{Massachusetts Bd. of Retirement v. Murgia}.
\newblock  \bibinfo{number}{{427 U.S. 307}} (\bibinfo{year}{1976}).
\newblock


\bibitem[Wang et~al\mbox{.}(2020)]%
        {wang2020factors}
\bibfield{author}{\bibinfo{person}{Ruotong Wang}, \bibinfo{person}{F~Maxwell Harper}, {and} \bibinfo{person}{Haiyi Zhu}.} \bibinfo{year}{2020}\natexlab{}.
\newblock \showarticletitle{Factors influencing perceived fairness in algorithmic decision-making: Algorithm outcomes, development procedures, and individual differences}. In \bibinfo{booktitle}{\emph{Proceedings of CHI 2020}}. \bibinfo{pages}{1--14}.
\newblock


\bibitem[Williams et~al\mbox{.}(2007)]%
        {williams2007understanding}
\bibfield{author}{\bibinfo{person}{James~Herbert Williams}, \bibinfo{person}{Richard~A Van~Dorn}, \bibinfo{person}{Charles~D Ayers}, \bibinfo{person}{Charlotte~L Bright}, \bibinfo{person}{Robert~D Abbott}, {and} \bibinfo{person}{J~David Hawkins}.} \bibinfo{year}{2007}\natexlab{}.
\newblock \showarticletitle{Understanding race and gender differences in delinquent acts and alcohol and marijuana use: A developmental analysis of initiation}.
\newblock \bibinfo{journal}{\emph{Social Work Research}} \bibinfo{volume}{31}, \bibinfo{number}{2} (\bibinfo{year}{2007}), \bibinfo{pages}{71--81}.
\newblock


\end{thebibliography}

\end{document}